# How We Tend To Overestimate Powerlaw Tail Exponents

**Nassim N. Taleb**
NYU Polytechnic Institute

October 2012

In the presence of a layer of metaprobabilities (from metadistribution of the parameters), the asymptotic tail exponent corresponds to the lowest *possible* tail exponent *regardless* of its probability. The problem explains "Black Swan" effects, i.e., why measurements tend to chronically underestimate tail contributions, rather than merely deliver imprecise but unbiased estimates.

## The Effect of Estimation Error, General Case

The idea of model error from missed uncertainty attending the parameters (another layer of randomness) is as follows. Most estimations in economics (and elsewhere) take, as input, an average or expected parameter, $\bar{\alpha} = \int \alpha\, \phi(\alpha)\, d\alpha$, where $\alpha$ is $\phi$ distributed ( deemed to be so a priori or from past samples), and regardles of the dispersion of $\alpha$, build a probability distribution for X that relies on the mean estimated parameter, $p(X) = p(X \mid \bar{\alpha})$, rather than the more appropriate metaprobability adjusted probability:

$$p(X) = \int p(X \mid \alpha)\, \phi(\alpha)\, d\alpha \tag{1}$$

In other words, if one is not certain about a parameter $\alpha$, there is an inescapable layer of stochasticity; such stochasticity raises the expected (metaprobability-adjusted) probability. The uncertainty is fundamentally epistemic, includes incertitude, in the sense of lack of certainty about the parameter.

The model bias becomes an equivalent of the Jensen gap (the difference between the two sides of Jensen's inequality), typically positive since probability is convex away from the center of the distribution. We get the bias $\omega_A$ [1-2] from the differences in the steps in integration

$$\omega_A = \int f(X)\, p(X \mid \alpha)\, \phi(\alpha)\, d\alpha \;-\; p\!\left(X \;\Big|\; \int \alpha\, \phi(\alpha)\, d\alpha\right) \tag{2}$$

With $f(X)$ a function , $f(X)=X$ for the mean, etc., we get the higher order bias $\omega_{A'}$

$$\omega_{A'} = \int\!\int f(X)\, p(X \mid \alpha)\, \phi(\alpha)\, d\alpha\, dX \;-\; \int f(X)\, p\!\left(X \;\Big|\; \int \alpha\, \phi(\alpha)\, d\alpha\right) dX \tag{3}$$

Now assume the distribution of $\alpha$ as discrete n states, with $\alpha = \{\alpha_i\}_{i=1}^n$ each with associated probability $\phi = \{\phi_i\}_{i=1}^n$ $\sum_{i=1}^n \phi_i = 1$. Then (1) becomes

$$p(X) = \sum_{i=1}^{n} p(X \mid \alpha_i)\, \phi_i \tag{4}$$

So far this holds for $\alpha$ any parameter of any distribution.

## Application to Powerlaws

Unlike the Gaussian; when the perturbation affects the standard deviation of a Gaussian, the end product is the weighted average probability distributions, a powerlaw distribution with errors about the possible tail exponent will bear the asymptotic properties of the *lowest* exponent, not the average exponent.

Now assume p(X) a standard Pareto Distribution with $\alpha$ the tail exponent being estimated, $p(X \mid \alpha) = \alpha\, X^{-\alpha-1}\, X_{\min}^{\alpha}$, where $X_{\min}$ is the lower bound for X,

$$p(X) = \sum_{i=1}^{n} \alpha_i X^{-\alpha_i-1} X_{min}^{\alpha_i} \phi_i \quad (5)$$

Taking it to the limit

$$\lim_{X \to \infty} X^{\alpha^*+1} \sum_{i=1}^{n} \alpha_i X^{-\alpha_i-1} X_{min}^{\alpha_i} \phi_i = K$$

where K a constant and $\alpha^* = \min_{1 \leq i \leq n} \alpha_i$. In other words $\sum_{i=1}^{n} \alpha_i X^{-\alpha_i-1} X_{min}^{\alpha_i} \phi_i$ is asymptotically equivalent to a constant times $X^{\alpha^*+1}$. The lowest parameter in the space of all possibilities becomes the dominant parameter for the tail exponent.

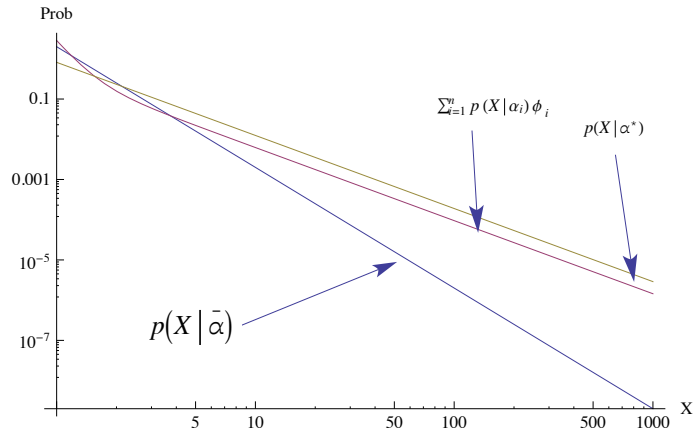

Figure 1: Log-log plot illustration of the asymptotic tail exponent with two states. The graphs shows the different situations, a) $p(X|\bar{\alpha})$ b) $\sum_{i=1}^{n} p(X|\alpha_i) \phi_i$ and c) $p(X|\alpha^*)$. We can see how b) and c) converge

The asymptotic Jensen Gap $\omega_A$ becomes $p(X|\alpha^*) - p(X|\bar{\alpha})$

## Implications

1. Whenever we estimate the tail exponent from samples, we are likely to underestimate the thickness of the tails, an observation made about Monte Carlo generated $\alpha$-stable variates and the estimated results [3].
2. The higher the estimation variance, the lower the true exponent.
3. The asymptotic exponent is the lowest possible one. It does not even require estimation.
4. Metaprobabilistically, if one isn't sure about the probability distribution, and there is a probability that the variable is unbounded and "could be" powerlaw distributed, then it is powerlaw distributed, and of the lowest exponent.

The obvious conclusion is to in the presence of powerlaw tails, focus on changing payoffs to clip tail exposures to limit $\omega_{A'}$ and "robustify" tail exposures, making the computation problem go away.[2]

## References

[1] Taleb, N.N, and Douady, R., 2012, Mathematical Definition and Detection of (Anti)Fragility.NYU-Poly

[2] Taleb, N.N, 2012, Antifragile, Random House and Penguin.

[3] Weron, R., 2001, Levy-stable distributions revisited: tail index > 2 does not exclude the Levy-stable regime. International Journal of Modern Physics C, 12(02):209–223.